\documentclass[epj,twocolumn]{webofc}
\usepackage[varg]{txfonts}   % Web of Conferences font
\woctitle{LHCP 2013}
\begin{document}
\title{Searches for New Physics in Multijet Final States}

\author{Carl Vuosalo\inst{1}
on behalf of the CMS Collaboration}

\institute{The Ohio State University
       }

\abstract{%
 A variety of new physics models predict heavy resonances that decay to multiple hadronic jets. These models include axigluons, colorons, diquarks, excited quarks, Randall-Sundrum gravitons, string resonances, and Z' models, among others. Other models make the prediction that high-$p_{T}$ jets will be suppressed, resulting in jet extinction. Using the data collected in 2012 at a center-of-mass energy of 8 TeV, the CMS Collaboration has made a baseline inclusive jet
 cross section 
 measurement for comparison with new-physics searches, and then performed searches for jet
extinction and resonances that decay to two hadronic jets. The results of these searches will be presented. No evidence of new physics has been observed, and these results set new limits on the parameters of these models.

}
\maketitle

\section{Introduction}
\label{sec:intro}

A major goal of the Compact Muon Solenoid (CMS) experiment~\cite{cmsdet} at the Large Hadron Collider (LHC)
is the discovery of new physics (NP) beyond the standard model (SM).
Many NP theoretical models predict jet extinction or
production of resonances that decay to paired jets. During 2012, the CMS
detector recorded 19.6~$\textrm{fb}^{-1}$ of pp data at a center-of-mass energy of 8 TeV.
Several CMS
analyses have exploited this rich data set to make a measurement of the inclusive jet cross section and to search for these NP signals.

\section{Measurement of Differential Inclusive Jet Cross Sections}
The inclusive jet cross section provides an important baseline for comparison with new-physics
models. Additionally, the measurement provides a method of assessment of Quantum Chromodynamic (QCD) calculation frameworks and parton distribution functions (PDFs) used with
those frameworks. CMS has performed the measurement of the double-differential inclusive jet
cross section in terms of $p_{T}$ and rapidity by using 10.7~$\textrm{fb}^{-1}$ of the 2012 8 TeV pp dataset~\cite{smp1212}. The measurement uses six, staggered triggers to cover a $p_{T}$ range
from 74--2500 GeV and a range of parton momentum fraction $x$ from 0.019--0.625. The data
are corrected for detector smearing effects and jet energy resolution. The dominant systematic
uncertainty for data comes from the jet energy scale and ranges from 15--40\%.

\begin{figure}
  \centering
  \includegraphics[width=8cm,clip]{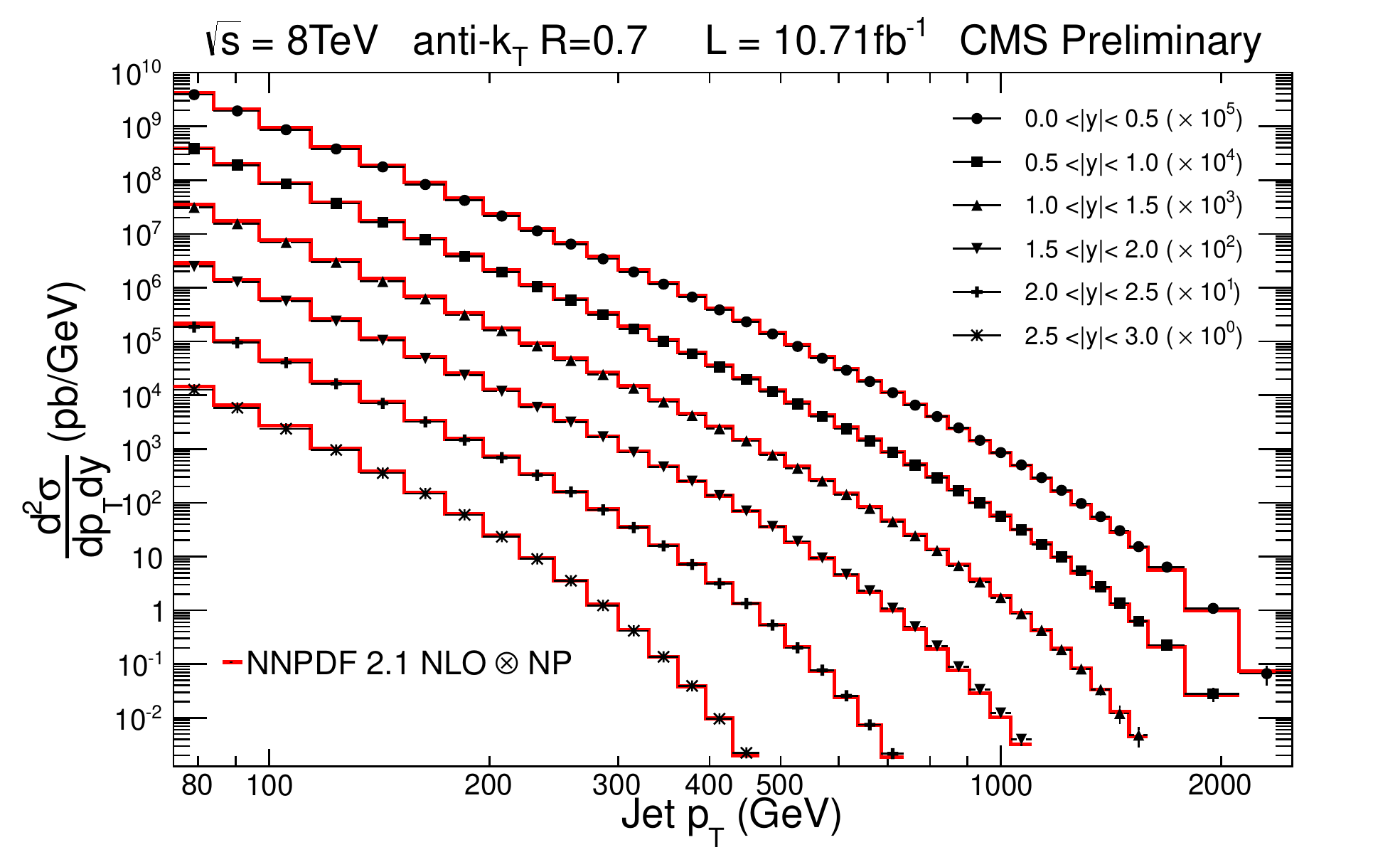}
  \caption{Double-differential inclusive jet cross section (points) in
    comparison to NLO predictions (red lines) using the NNPDF2.1 PDF set.
    Data points are shown for six rapidity ranges.}
  \label{fig:ResultNNPDF}
\end{figure}

The theory prediction is calculated at next-to-leading-order (NLO)
by NLOJet++~\cite{Nagy:2001fj, Nagy:2003tz}
in the fastNLO framework~\cite{Britzger:2012bs}. Its dominant  systematic uncertainty comes from PDF variation and ranges from 10--50\%. The measured cross section matches the theory calculation with the NNPDF2.1 PDF set~\cite{NNPDF},
as shown in Fig.~\ref{fig:ResultNNPDF}.

Data was compared to predictions from five PDF sets:
ABM11~\cite{ABKM11}, CT10~\cite{CT10}, HERAPDF1.5~\cite{HERA},
MSTW2008NLO~\cite{MSTW},
and NNPDF2.1. All match data within the uncertainties except for ABM11 in certain rapidity ranges.
These results provide strong confirmation for the NLOJet++ calculations and PDFs.

\section{Search for Jet Extinction in the Inclusive Jet $p_{T}$ Spectrum}
Terascale gravity theories predict that the Planck scale would occur at the TeV scale~\cite{1999Banks}. In this
case, collisions at the LHC could produce microscopic black holes. Many such theories predict
that these black holes would decay spectacularly in a
spray of high-$p_{T}$ jets and particles~\cite{ref:Giddings_and_Thomas,ref:Landsberg_Dimopoulos}, but
LHC searches so far have found no evidence for such black hole
decay~\cite{StrongGravity_ATLAS2012,Diphoton_ATLAS2012,ZZ_ATLAS2012,
Black_Holes_CMS2012,Diphotons_CMS2012,Dijet_CMS2011,Dilepton_CMS2011,CMS_monophotons,
CMS_monojets}. 
However, non-perturbative
processes could produce black holes that decay to a high multiplicity of
low-energy jets~\cite{Kilic:2012wp}. Such
black hole decays might be difficult to distinguish from the low-energy SM background, but their
effect on the jet-$p_{T}$ spectrum would be dramatic. At a certain energy scale, black
holes would dominate and thereby suppress the production of high-$p_{T}$ jets, causing
jet extinction at the high end of the jet-$p_{T}$ spectrum.

\begin{figure}
\centering
\includegraphics[width=8cm,clip]{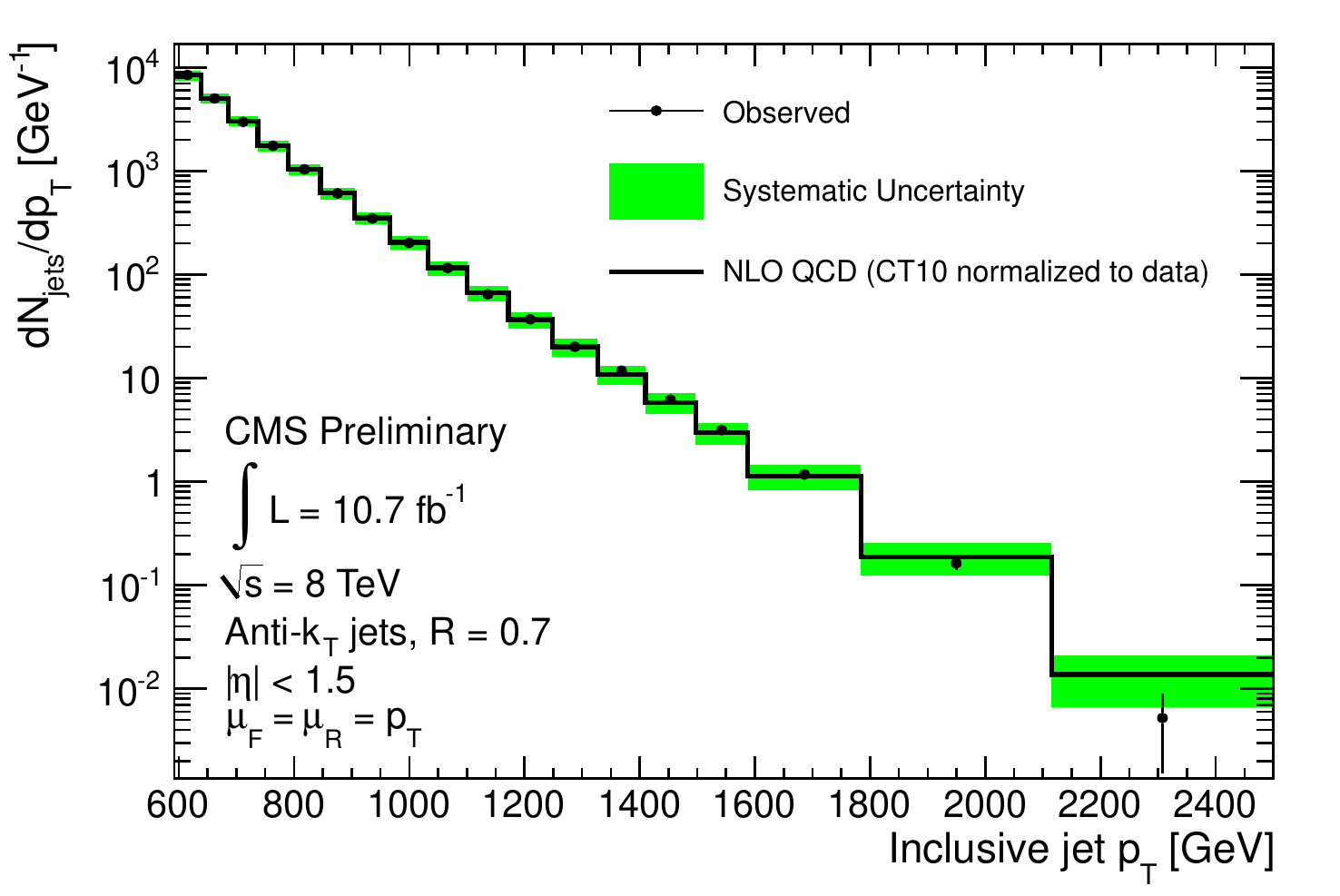} 
\caption{Inclusive jet $p_{T}$ spectrum (points) for $|\eta| < 1.5$, as observed in
10.7~$\textrm{fb}^{-1}$ of data.   The SM NLO 
simulation, convolved with the detector response and
normalized to the total observed cross section, is shown by the solid line.  The colored band shows the magnitude of all sources of systematic uncertainty added in quadrature.}
\label{fig:inclusive_jet_pt}
\end{figure}

\begin{figure}[h]
\centering
\includegraphics[width=8cm,clip]{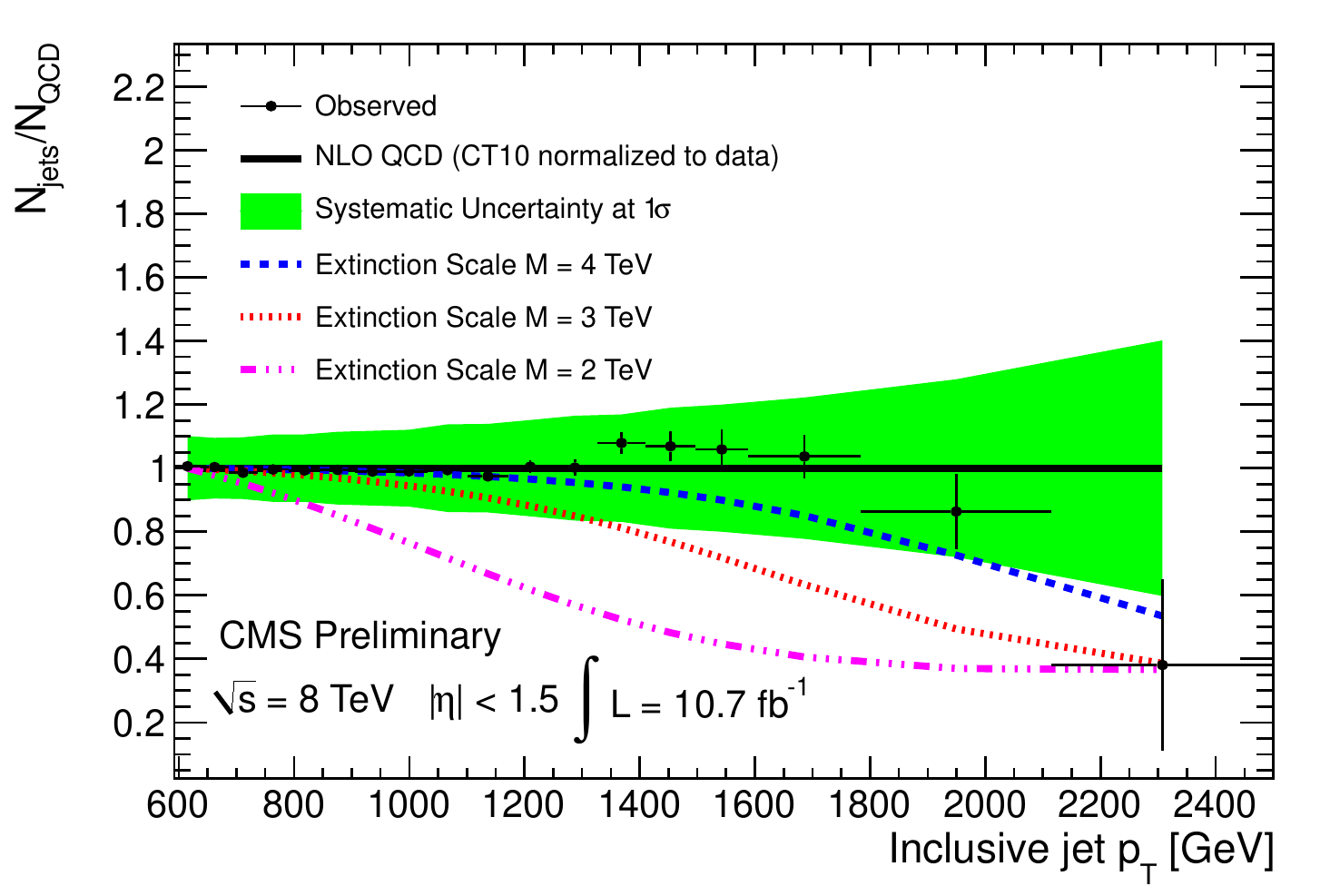} 
\caption{The ratio of the inclusive jet-$p_{T}$ spectrum to the NLO QCD prediction convolved with the detector resolution.  The colored band shows the magnitude of all sources of systematic uncertainty added in quadrature at 1$\sigma$.  Dashed lines indicate the effects of extinction at three different values of the extinction scale, $M=2$, 3, and 4~TeV.  
}\label{fig:jet_pt_ratios}
\end{figure}

\begin{figure}[h]
\centering
\includegraphics[width=8cm,clip]{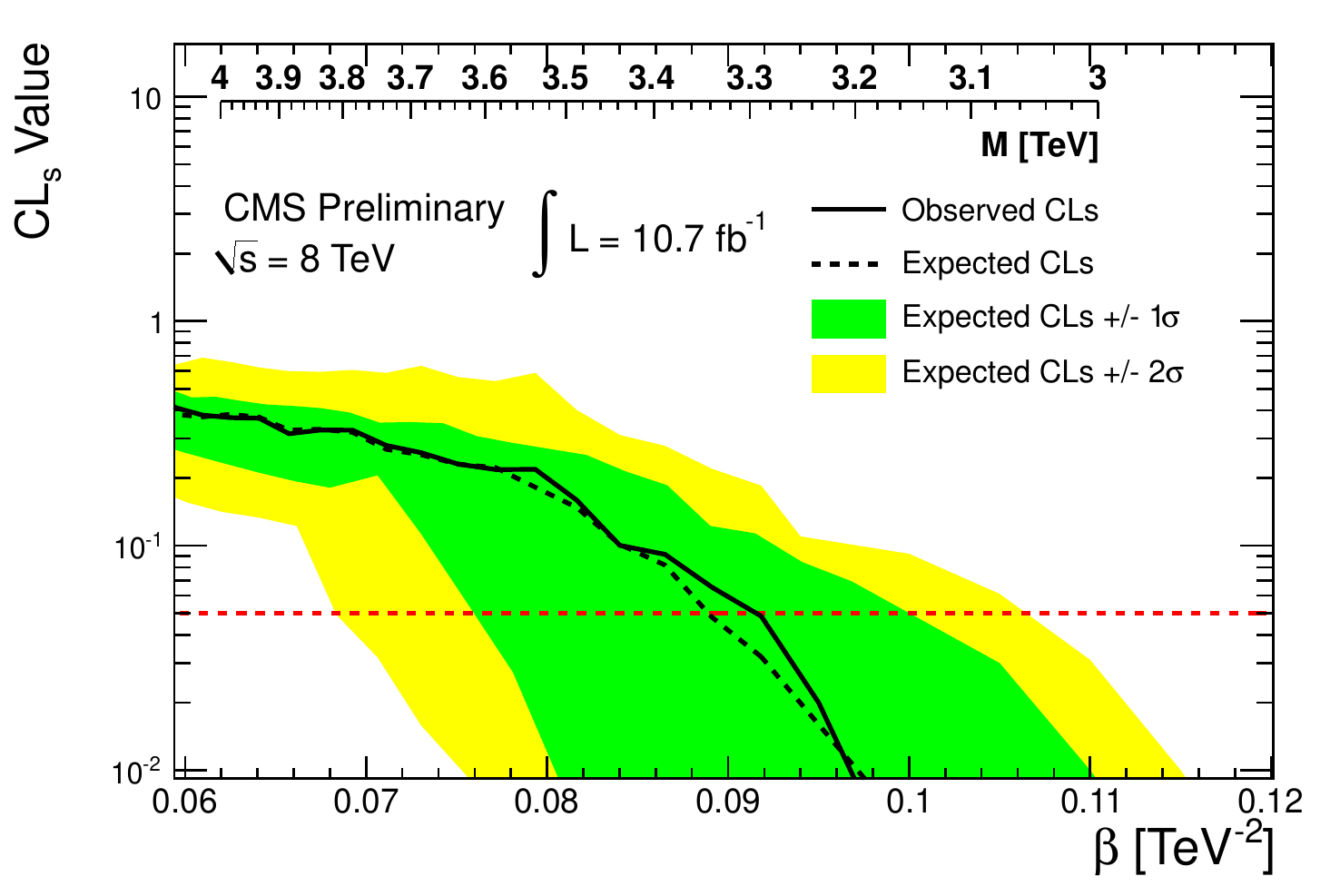}
\caption{The results of a ${\rm CL_{s}}$ scan in the extinction scale,
$\beta = M^{-2}$.    
The observed value of ${\rm CL_{s}}$ as a function of $\beta$ is shown by the solid line.  The observed upper limit on $\beta$ is $0.091~TeV^{-2}$ at 95\% CL, corresponding to a lower limit of 3.3~TeV on the extinction scale $M$.  The dashed line indicates the expected median of results for the background-only
hypothesis.  The green (dark) and yellow (light) bands indicate the ranges that are
expected to contain 68\% and 95\% of all observed excursions of the background from the median, respectively. The red line represents the 95\% CL.
}\label{fig:CLSLimit}
\end{figure}

CMS has performed a search for such jet extinction with 10.7~$\textrm{fb}^{-1}$ of the 2012 8 TeV pp dataset~\cite{exo1251}. The SM prediction was made at NLO with NLOJet++ in the fastNLO
framework and the CT10 PDF set and was scaled to data. The leading-order extinction spectrum
was generated with PYTHIA~\cite{ref:Pythia} and assumed the strong-coupling limit of the string model, so jet
extinction would occur beyond a scale $M$. The largest systematic uncertainty comes from
the jet energy scale and is about 10\%. The data was found to match the  SM prediction within
the uncertainties, as seen in Figs.~\ref{fig:inclusive_jet_pt} and ~\ref{fig:jet_pt_ratios}.
A  95\% confidence level (CL)
limit at 3.3 TeV on scale $M$ was set with the $\textrm{CL}_{\textrm{S}}$ calculator, as shown in Fig.~\ref{fig:CLSLimit}.

For the NLOJet++ calculator, there are a number of PDF sets that can be employed,
in addition to the CT10 PDF used for the SM estimate. The variation between PDF sets is
bracketed by the CT10 and MSWT2008 sets. Thus, as a cross check, the MSWT2008 set was tried
for the SM estimate, and it increased the observed limit by only 10\%, thereby confirming the
CT10 limit of 3.3 TeV as a conservative value.

\begin{figure*}[htbp]
  \begin{center}
    \includegraphics[width=7cm,clip]{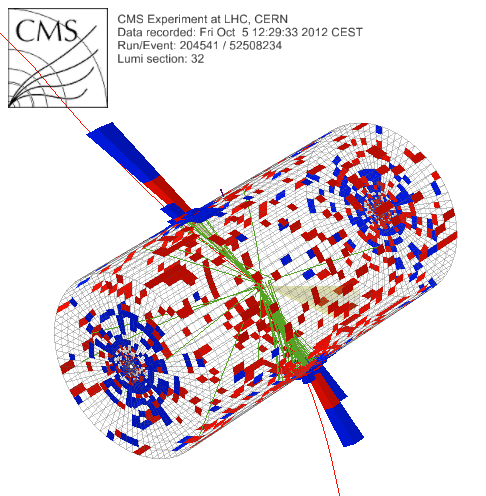}
    \includegraphics[width=7cm,clip]{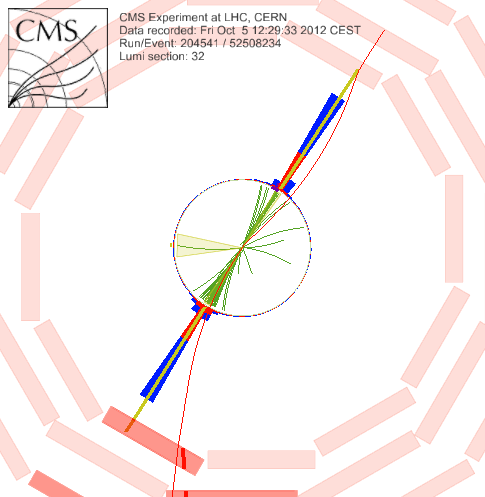}
    \caption{The event from the dijet search with the highest invariant mass: 3D view (left) and 2D view (right). The invariant mass of the two wide jets is 5.15 TeV.}
    \label{figEvent}
  \end{center}
\end{figure*}

\begin{figure}[htbp]
  \begin{center}
    \includegraphics[width=8cm,clip]{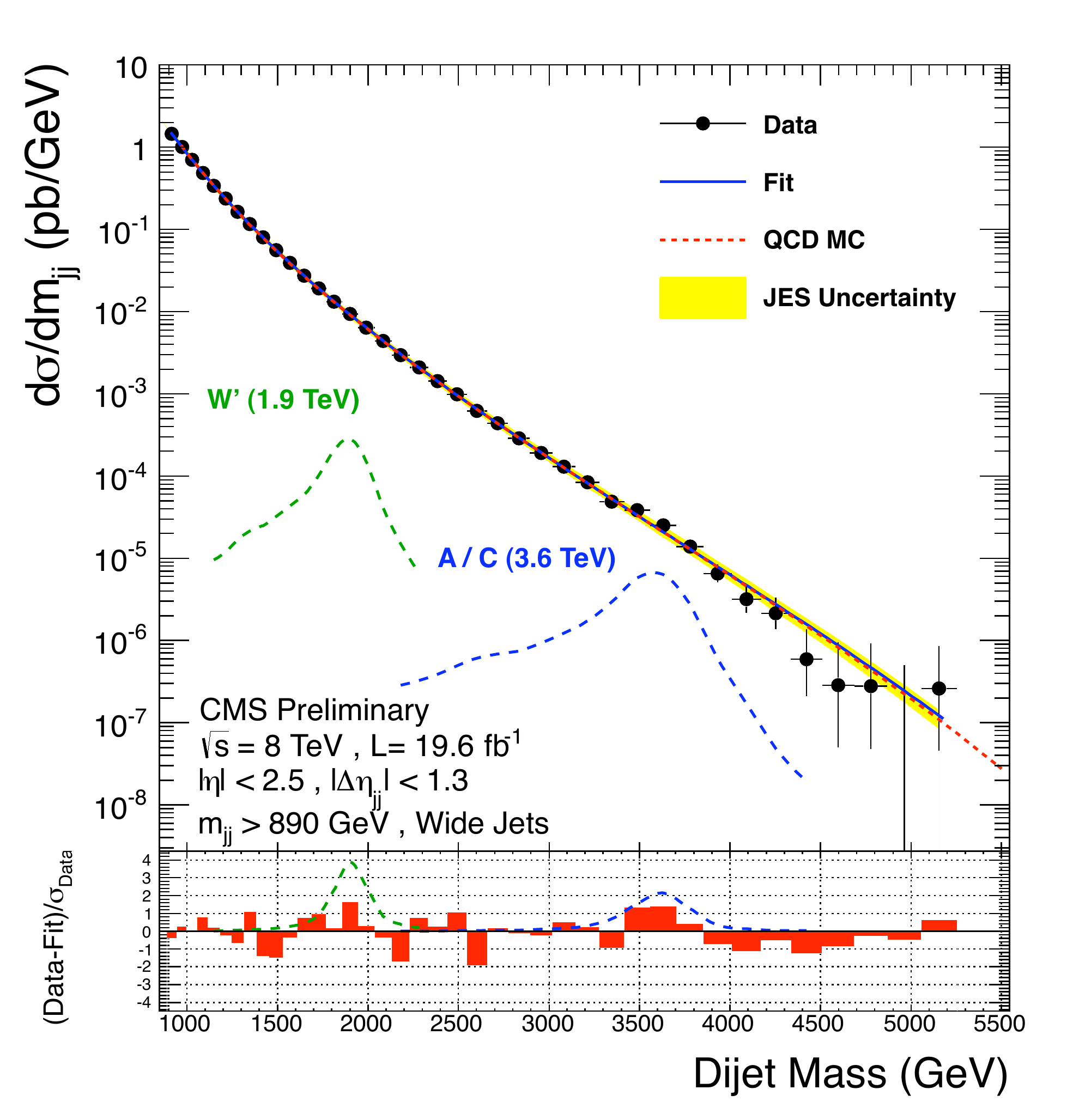}
    \caption{Dijet mass spectrum from wide jets (points) compared to a smooth fit (solid) and to 
predictions including detector simulation of QCD and signal resonances. 
The QCD prediction has been normalized to the data. 
The error bars are statistical only. 
The bin-by-bin fit residuals, $\mbox{(data-fit)}/\sigma_{\rm{data}}$, are shown at the bottom.
}
    \label{figDataAndMC}
  \end{center}
\end{figure}

\begin{figure}[!htp]
  \begin{center}
    \includegraphics[width=8cm,clip]{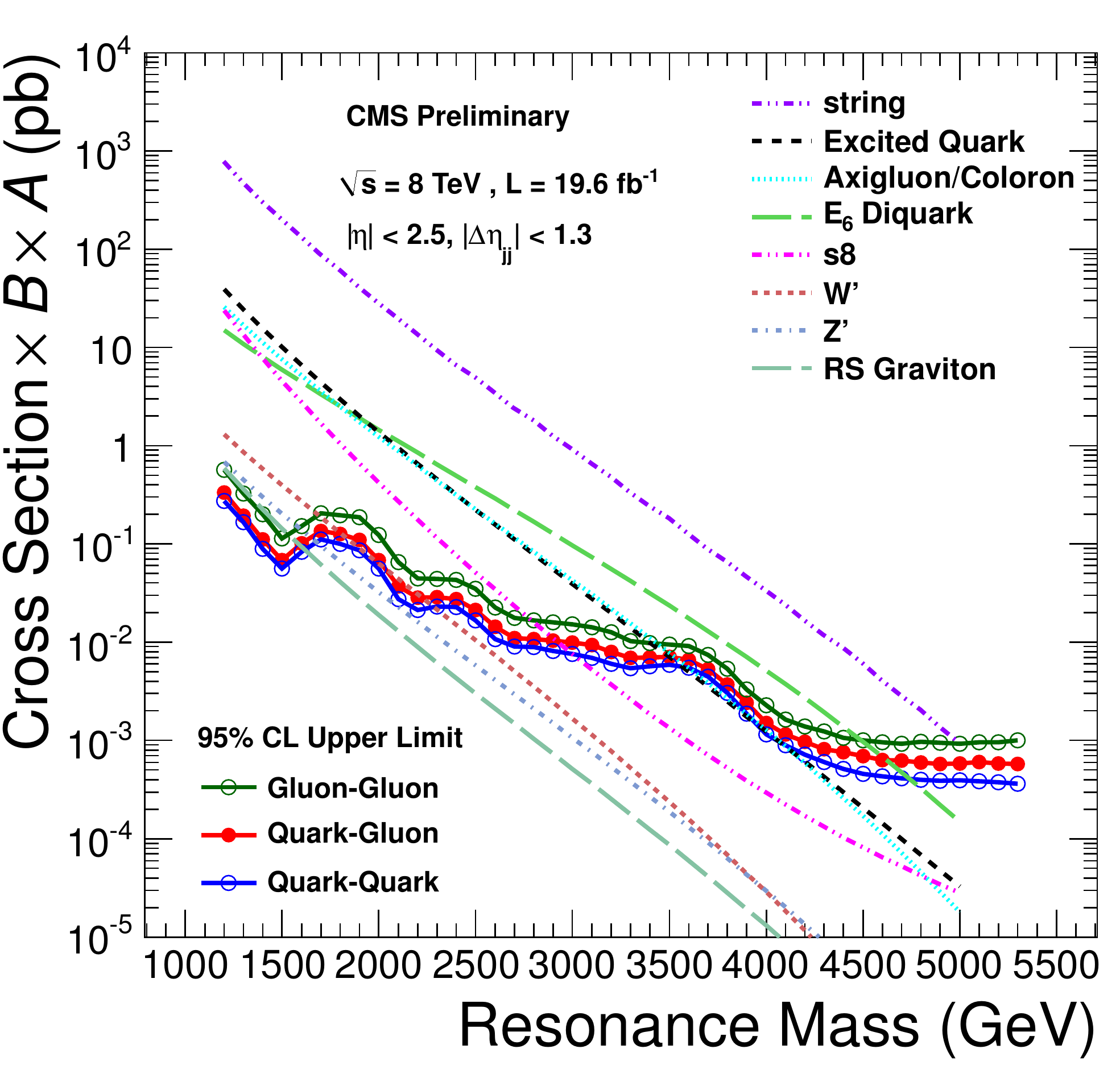}
    \caption{The observed 95\% CL upper limits on $\sigma\times B\times A$
for dijet resonances of the type gluon-gluon, quark-gluon, and quark-quark,
compared to theoretical predictions for string resonances, $\mbox{E}_6$ diquarks,
excited quarks, axigluons,
colorons, s8 resonances, 
new gauge bosons $\mbox{W}^{\prime}$ and $\mbox{Z}^{\prime}$,
and RS gravitons. 
}
    \label{figLimit}
  \end{center}
\end{figure}

\begin{figure}[!htp]
  \begin{center}
    \includegraphics[width=8cm,clip]{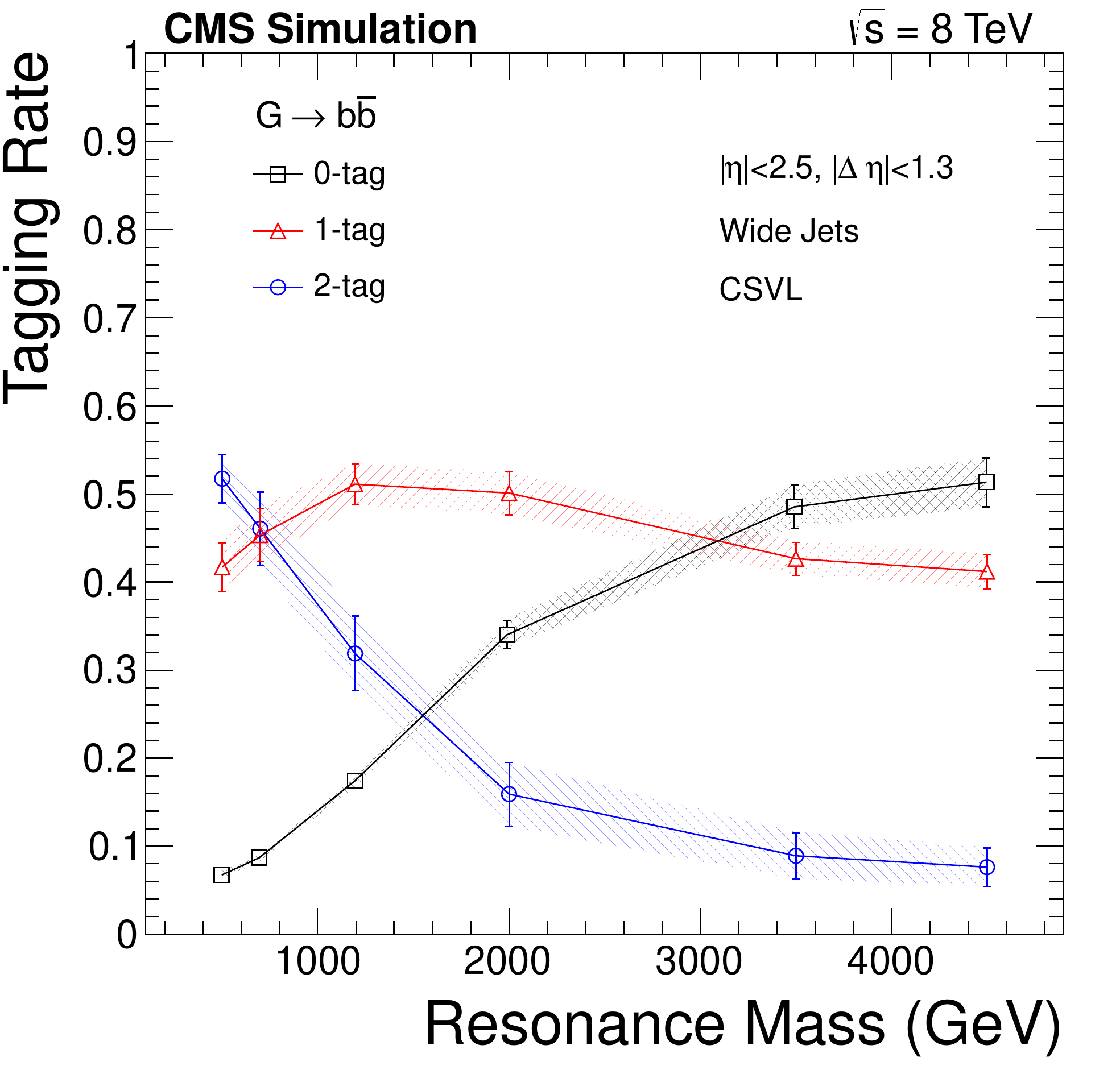}
    \caption{Tagging rates for 0, 1, and 2 b tags as a function of the resonance mass for the b$\overline{\textbf{b}}$ decay mode.
The hatched regions represent uncertainties in the tagging rates due to the variation of the b-tag scale factors within their uncertainties.}
    \label{fig:rate}
  \end{center}
\end{figure}

\section{Search for Narrow Resonances using the Dijet Mass Spectrum}
\label{sec:dijet}
Many new physics models predict heavy resonances that couple to quarks and gluons and decay
to dijets. Some of these models include axigluons (A)~\cite{ref_axi,Chivukula:2011ng},
color-octet colorons (C)~\cite{ref_coloron}, excited quarks~\cite{ref_qstar,Baur:1989kv}, 
Randall-Sundrum gravitons~\cite{ref_rsg}, scalar diquarks~\cite{ref_diquark},
string resonances~\cite{Anchordoqui:2008di,Cullen:2000ef},
technicolor s8 resonances~\cite{Han:2010rf}, and new gauge bosons (W' and Z')~\cite{ref_gauge}.
CMS has performed searches for such resonances, first with 4.0~$\textrm{fb}^{-1}$~\cite{exo1216},
and then with the full 19.6~$\textrm{fb}^{-1}$ 2012 8 TeV pp dataset~\cite{exo1259}. This search employs the wide-jet
technique~\cite{Cacciari:2008gd,Krohn:2009th,Abdesselam:2010pt}, which adds close sub-leading
jets to the two leading jets in each selected event. Figure~\ref{figEvent} shows a CMS
event display of the dijet event with the highest invariant mass in this dijet search.

The background prediction comes from a
four-parameter fit to the data. The largest systematic uncertainty is the
jet energy resolution uncertainty, which is 10\%. The data matches the background estimate with
no excess or bumps observed on the smooth background, as shown in Fig.~\ref{figDataAndMC}. Mass limits are set on the eight signal
models, as shown in Fig.~\ref{figLimit}, with the strongest limit being 5.1 TeV on the string
resonance mass.

\section{Search for Heavy Resonances Decaying into b$\overline{\textbf{b}}$ and bg Final States}
As a variation on the dijet search in Sec.~\ref{sec:dijet}, a b-jet tagging requirement can be
placed on the jets in order to reduce SM backgrounds and to make the search
sensitive to models that specifically produce b jets: excited b quarks~\cite{ref_qstar,Baur:1989kv}, RS gravitons~\cite{ref_rsg}, and a
sequential SM Z'~\cite{ref_gauge}. CMS has performed such a search with the full 19.6~$\textrm{fb}^{-1}$ 2012 8 TeV pp dataset~\cite{exo1223}. Like the previous analysis, this one uses the wide-jet technique and
a background estimate from a four-parameter fit to the data.

Because the b-tagging rate drops significantly for high-mass signal resonances, as shown in
Fig.~\ref{fig:rate}, the analysis is performed in three channels: 0 b tags, 1 b tag, and
2 b tags.

\begin{figure*}[btp]
  \begin{center}
    \includegraphics[width=0.32\textwidth]{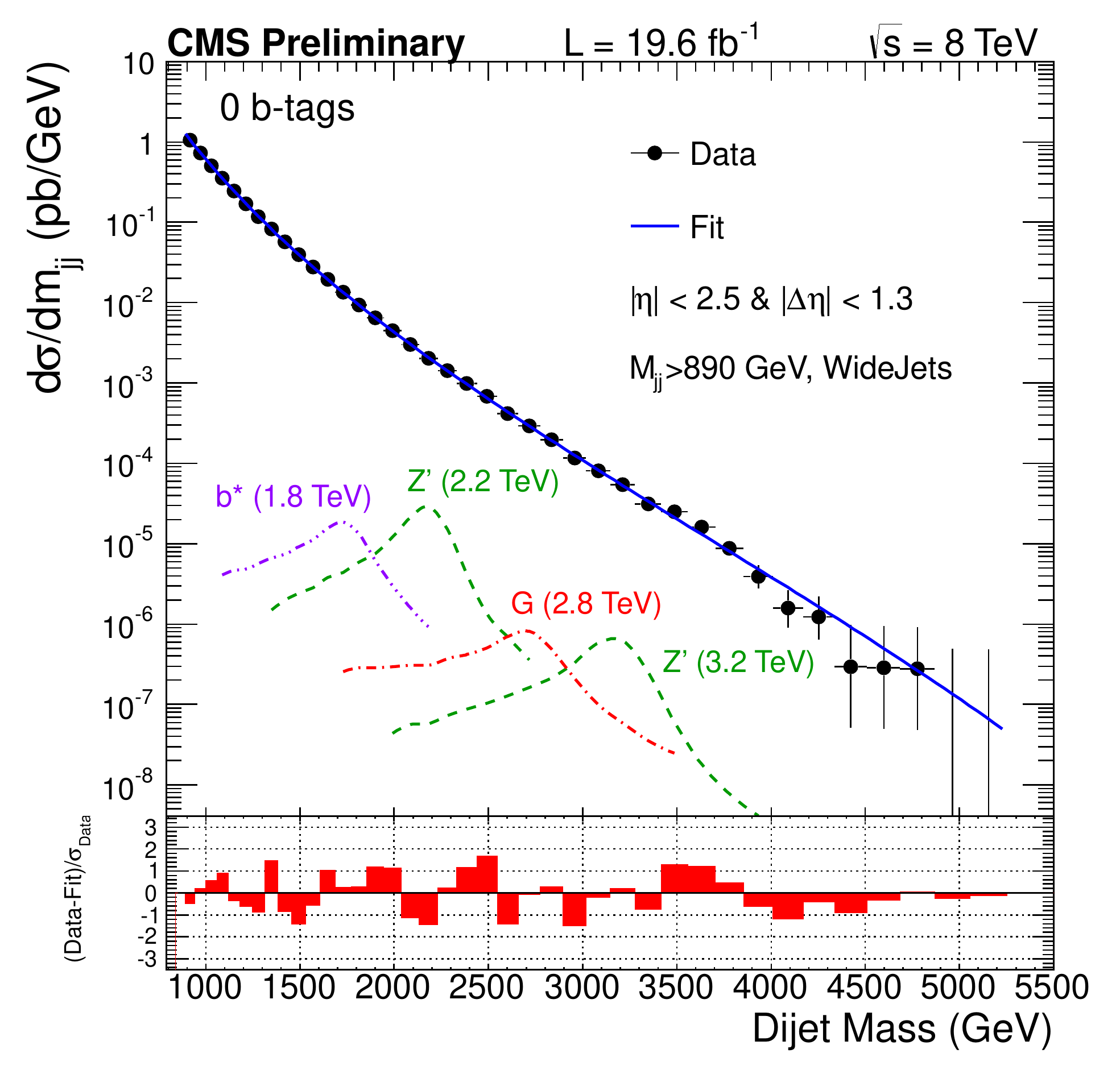}
    \includegraphics[width=0.32\textwidth]{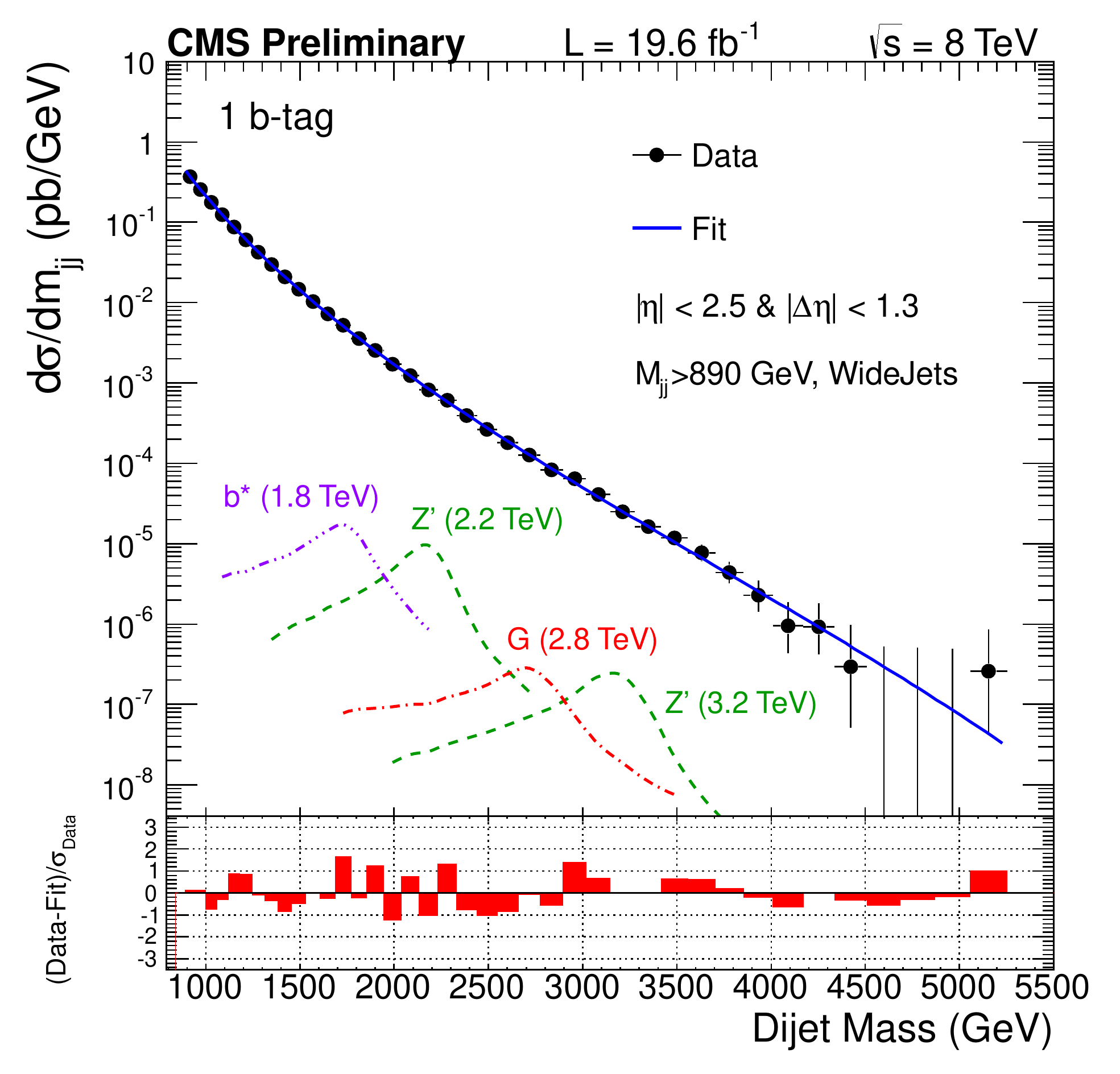}
    \includegraphics[width=0.32\textwidth]{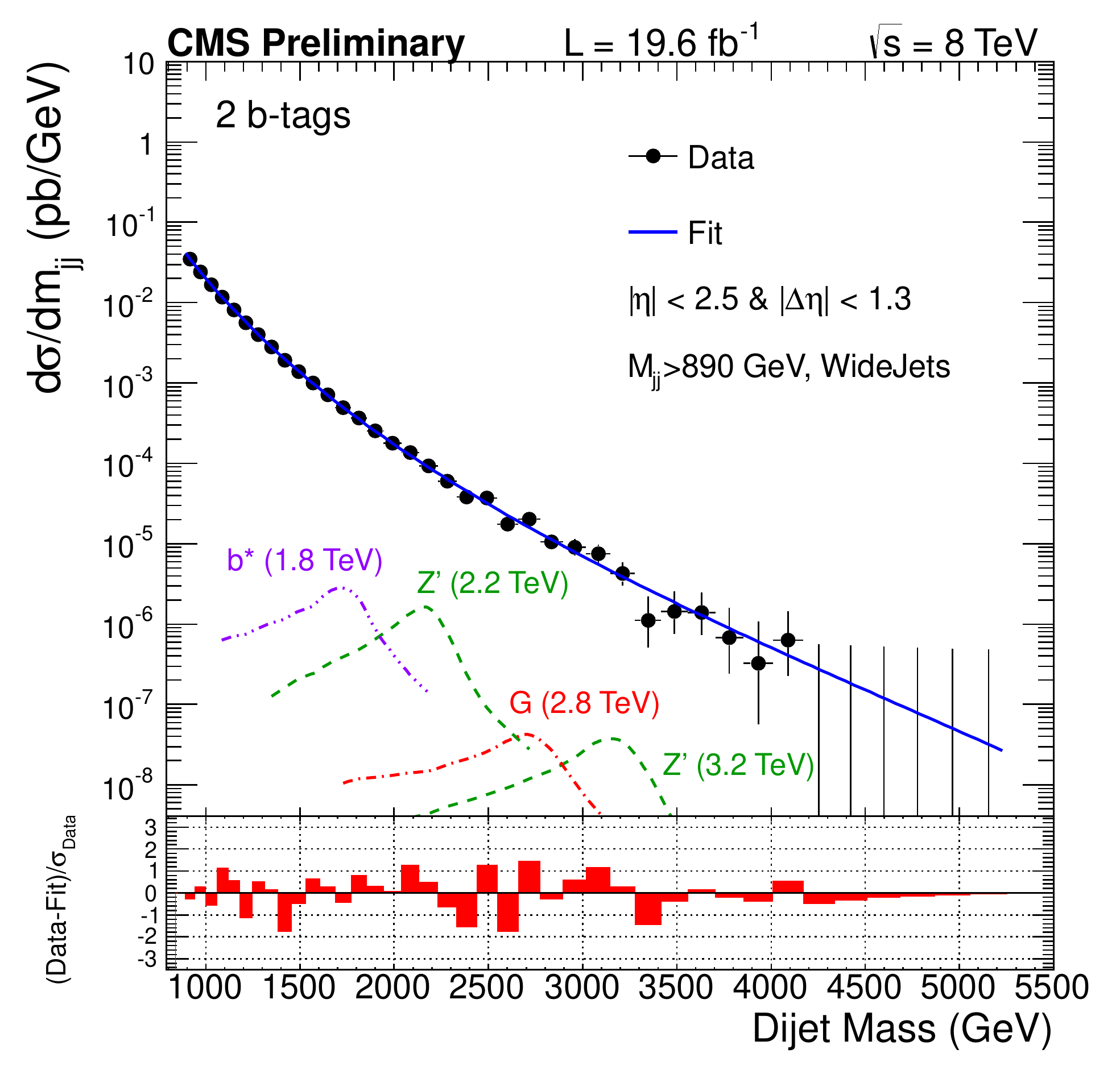}
    \caption{Dijet mass spectra (points) in different b-tag multiplicity bins compared to a smooth
fit (solid line). The bin-by-bin fit residuals are shown at the bottom of each plot. Predictions for RS graviton, Z', and excited b-quark signal spectra are also shown.
The vertical error bars are central intervals with correct coverage for Poisson variation, and the horizontal error bars are the bin widths.
}
    \label{fig:fit}
  \end{center}
\end{figure*}

\begin{figure*}[btp]
  \begin{center}
    \includegraphics[width=0.32\textwidth]{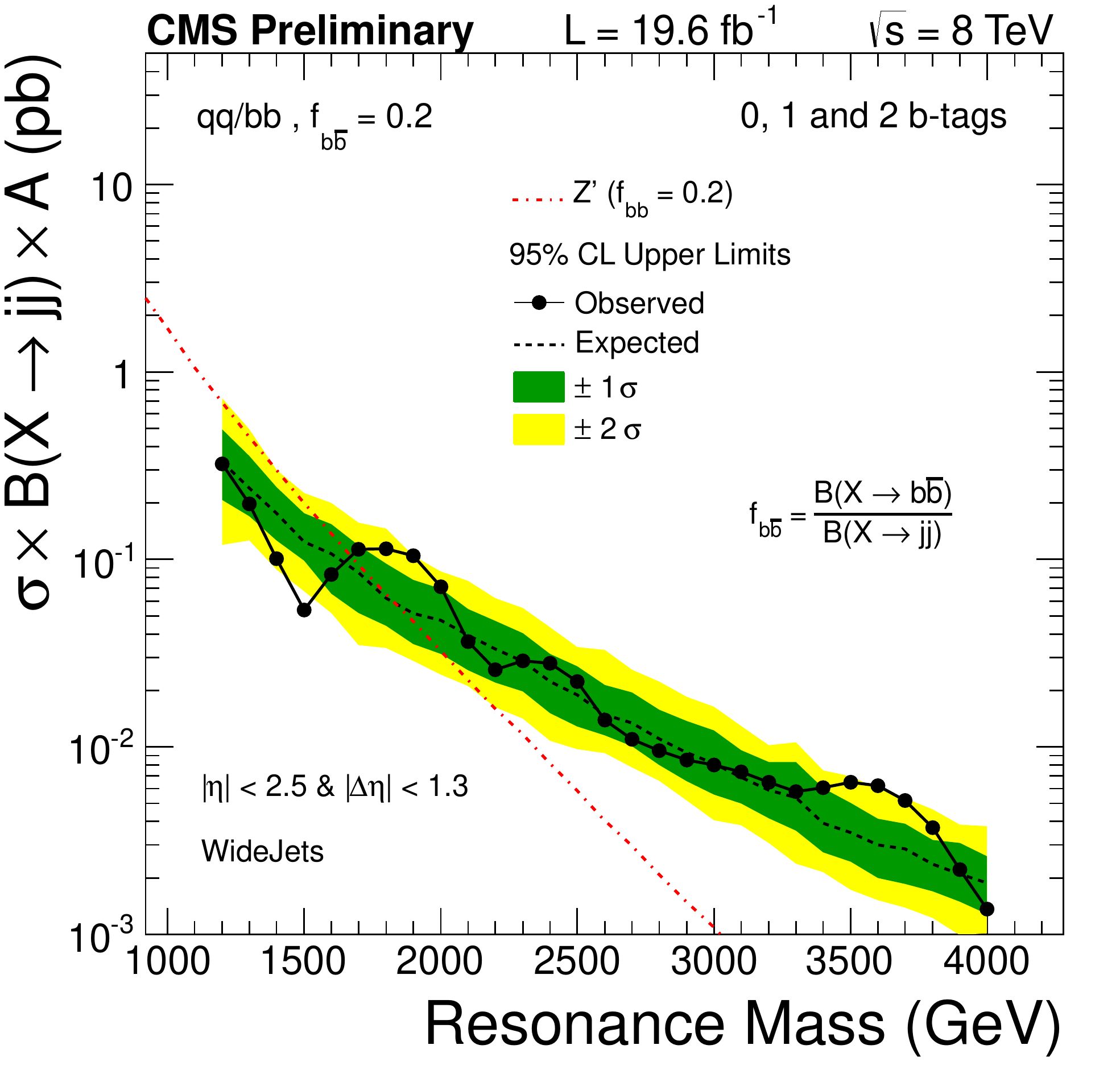}
    \includegraphics[width=0.32\textwidth]{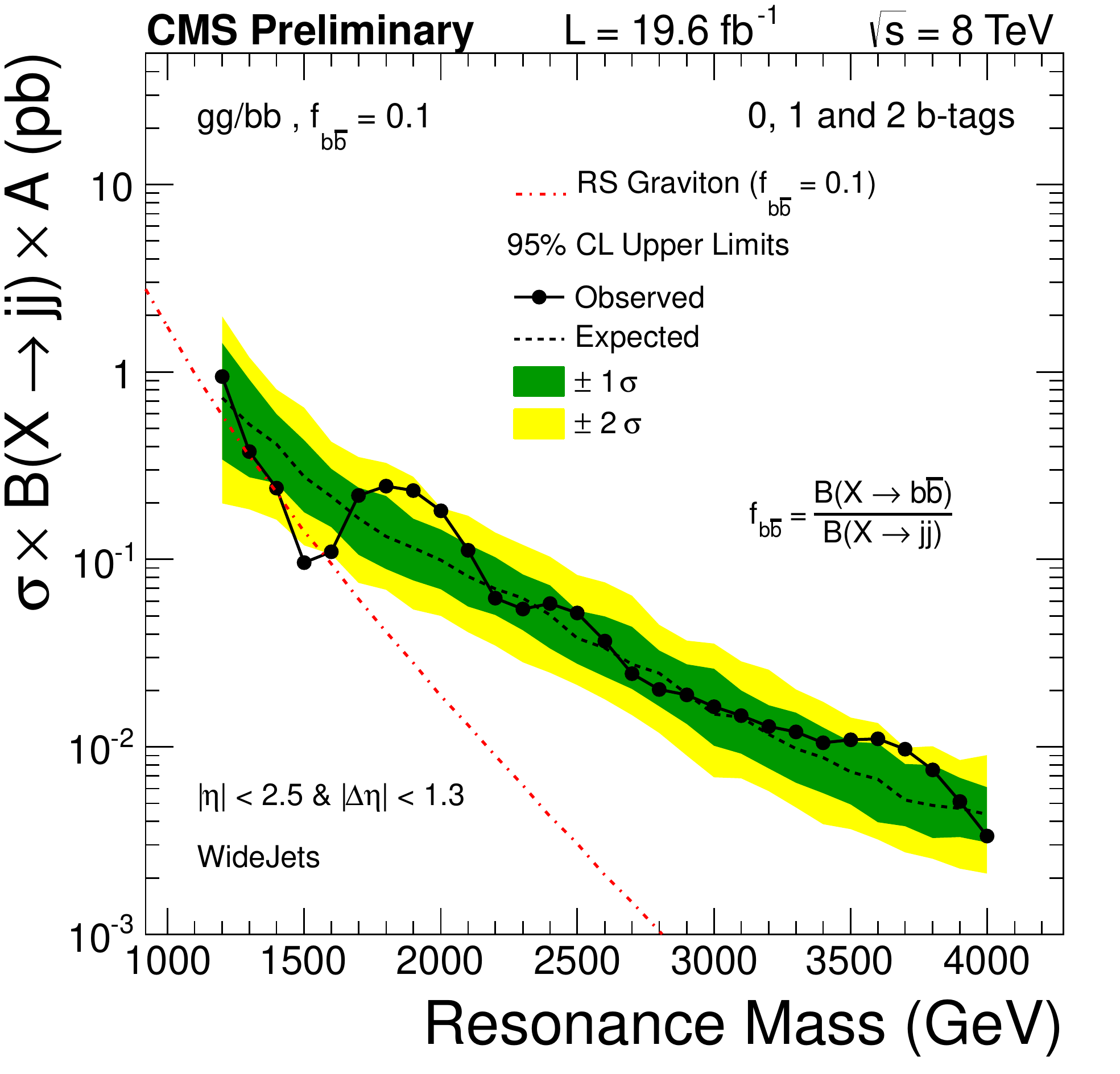}
    \includegraphics[width=0.32\textwidth]{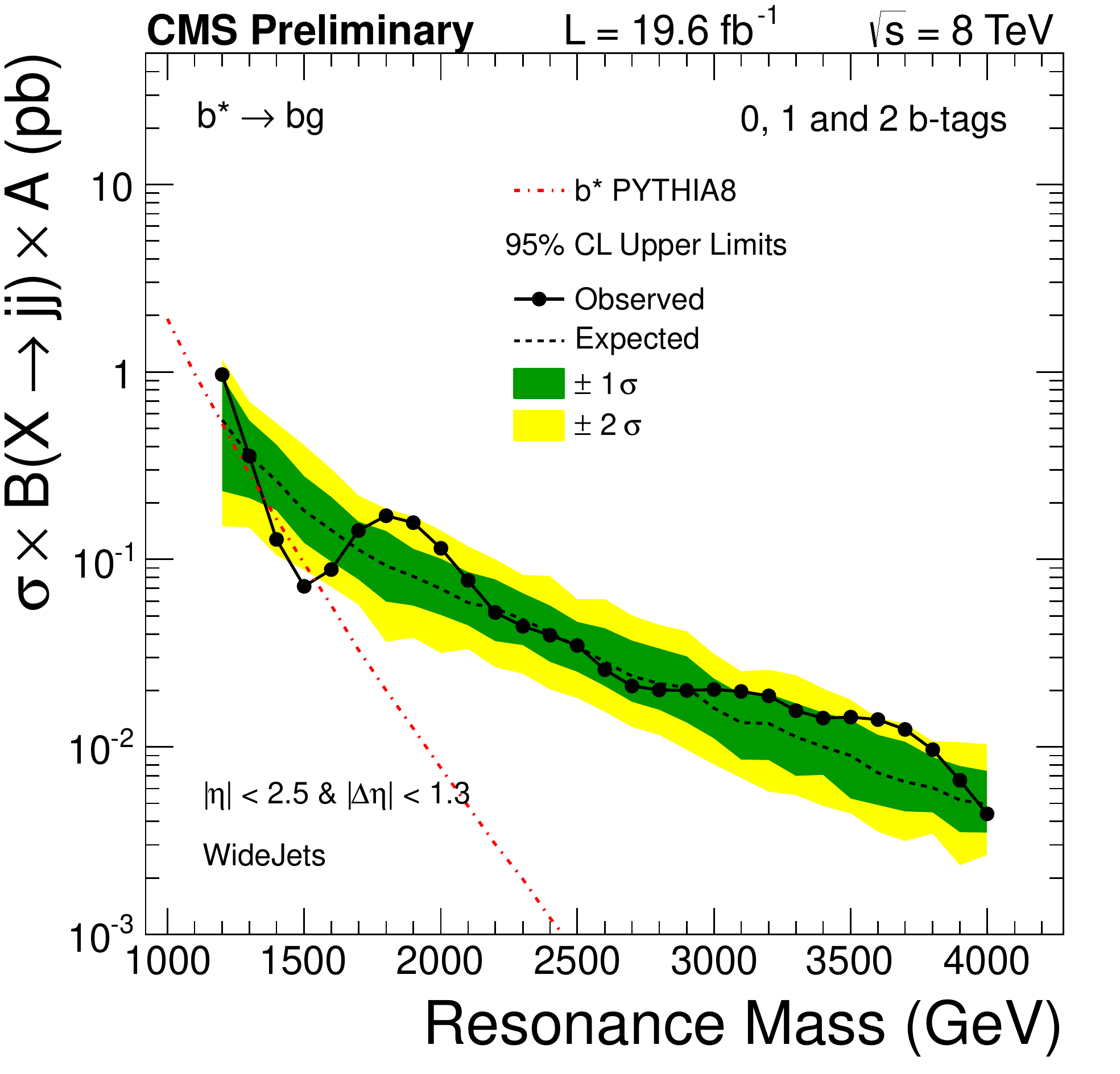}
    \caption{Combined observed and expected 95\% CL upper limits on $\sigma \times B \times A$ with systematic uncertainties
included for qq/bb resonances with $f_{\textrm{b} \bar{\textrm{b}}} = 0.2$ (left), gg/bb resonances with $f_{\textrm{b} \bar{\textrm{b}}} = 0.1$ (middle), and $\textrm{b}^{*} \rightarrow$ bg resonances (right). $f_{\textrm{b} \bar{\textrm{b}}}$ is the ratio
of the branching fraction of the resonance decaying to $\textrm{b} \bar{\textrm{b}}$ over the resonance decaying to all jets.
Theoretical cross sections for RS graviton, Z', and excited b quark are shown for comparison.}
    \label{fig:limits_obs_exp}
  \end{center}
\end{figure*}

The largest systematic uncertainty comes from the jet energy resolution and is about 10\%. The
background estimates compared to the data are shown in Fig.~\ref{fig:fit}.
The data match the  background estimates within the uncertainties, and no excess is observed. The best mass limits to date are set on the three signal models, as shown in Fig.~\ref{fig:limits_obs_exp}, with the strongest limit being 1.7 TeV on the Z'
mass.

\section{Conclusion}
The CMS Collaboration has  measured the inclusive jet cross section and 
performed several searches for new physics using jets
with the 2012 8 TeV pp collision dataset. The jet cross section measurement has confirmed
the NLOJet++ calculations and PDF sets.
In the searches, no significant deviations from the SM were observed.
New limits have been set on many models, most being the best to date on these models. These limits
range up to 3.3 TeV on the jet extinction scale,
up to 5.1 TeV on dijet string resonances, and up to 1.7 TeV on a Z'. CMS is continuing its searches
for new physics, and more results will be coming out soon.

%For one-column wide figures use syntax of figure~\ref{fig-1}

%
% BibTeX or Biber users please use (the style is already called in the class, ensure that the "woc.bst" style is in your local directory)
\bibliography{LHCP2013-Vuosalo}

\end{document}